\documentclass[aps,prl,showpacs,twocolumn,groupedaddress,amssymb]{revtex4}
\usepackage{graphicx}% Include figure files
\usepackage{dcolumn}% Align table columns on decimal point
\usepackage{bm}% bold math

\begin{document}

%\preprint{1st paper}

\title{Overcoming the Child-Langmuir law via the magnetic mirror effect}

\author{S. Son}
\affiliation{18 Caleb Lane, Princeton, NJ 08540}
\author{Sung Joon Moon}
\affiliation{28 Benjamin Rush Lane, Princeton, NJ 08540}
\date{\today}

\begin{abstract}
The maximum current in a vacuum tube prescribed by the classical Child-Langmuir law
can be overcome, when the space-charge effect of the induced potential is
mitigated by the mirror effect in a spatially varying magnetic field.
The current could exceed the Child-Langmuir value by as much as a few factors.   
The regime of practical interest is examined.  
\end{abstract}

\pacs{41.20.-q,  52.38.-r,  42.60.-v}                   % Classification Scheme.
\maketitle

%The maximum electric current in a vacuum tube is prescribed by
The Child-Langmuir law gives the one-dimensional maximum space-charge-limited
current in a vacuum tube~\cite{child, lang, qchild, qchild2, child3, child4}.
The steady state in a vacuum tube requires that the static potential energy
at a given point between the electrodes does not exceed that of the cathode.
In the following, we refer to this requirement as the {\it non-negativeness},
which is necessary because the electrons emitted from the cathode, 
usually of zero kinetic energy, need to reach the anode. 
However, at an intense enough current,
the potential no longer maintains this feature 
as a steep curvature in the static potential gets induced 
by the space-charge effect, 
which is the origin of the Child-Langmuir law.

In this paper,
the effect of a spatially varying magnetic field inside a vacuum tube
is investigated, in the context of the Child-Langmuir law. 
Let us assume there exists an magnetic field, of which
strength decreases from the cathode to the anode. 
If the electron emitted from the cathode initially has 
some perpendicular kinetic energy, 
the electron would gain the parallel kinetic energy
as it moves toward the anode due to the magnetic moment conservation,
or the magnetic mirror effect~\cite{mirror, mirror2}.
The goal of this paper is to demonstrate that this effect eventually leads to
overcome the Child-Langmuir law. 
In other words, the electrons would reach the anode
even when the self-induced static potential
does not satisfy the non-negativeness.
In usual circumstances, 
the electrons emitted from the cathode do not have 
the perpendicular kinetic energy, however, for example, 
the energy could be injected into the electrons at the cathode 
by a microwave E\&M wave via the cyclotron resonance.
We estimate the physical parameters 
in which this effect could be of practical interest.

One-dimensional fluid equations for electrons in a vacuum tube,
together with the Poisson equation, are
\begin{eqnarray} 
\frac{ \partial n_e }{ \partial t }  + \frac{\partial (n_e v_x)}{ \partial x} &=& 0   \mathrm{,}\nonumber \\ \nonumber \\
\frac{ \partial v_x }{ \partial t }  +  v_z\frac{\partial v_z}{ \partial x} 
&=& \frac{e}{m_e} \frac{ \partial \phi }{ \partial x } \mathrm{,} \nonumber \\ \nonumber \\
\frac{ \partial^2 \phi }{ \partial x^2 } &=& 4\pi e n_e \nonumber \mathrm{,} \\ \nonumber  
\end{eqnarray}
where $n_e = n_e(x)$ is the electron density between the cathode ($x=0$) and the anode ($x=d$), 
$v_x = v_x(x)$ is the electron velocity, and $\phi = \phi(x) $ is the self-induced static potential.  
The boundary conditions are given as $ v_x(0) = 0 $, $\phi(0) = 0 $ and $\phi(d) = V_0 $, 
where the cathode and the anode
% separated by $d$
has the bias of $V_0$.
At the steady state, the first two equations are simplified to 
$n_e(x) v_x(x) = {C_1} $ and $1/2 m_e v_x^2(x) - e\phi(x) = {C_2}$,
where $C_1$ and $C_2$ are constants.
%in addition to the Poisson equation. 
Then the solution can be simplified to 
\begin{equation} 
 \frac{ \partial^2 \phi(x) }{ \partial x^2  } =
\frac{4\pi J}{ \sqrt{ \frac{2e}{m_e} \phi(x)}} \mathrm{,}
\label{eq:sta}
\end{equation}
where $J = e n v_x(x) $ is the current density. 
The maximum current maintaining the non-negativeness is
\begin{equation}
  J_{\mathrm{max}} = \frac{4}{9} \left(\frac{2e}{m_e}\right)^{1/2} \frac{1}{ 4 \pi} 
  \frac{V_0^{3/2}}{d^2} \mathrm{,}
\label{eq:child}
\end{equation}
which is the Child-Langmuir law. 

Let us  denote  the magnetic field in the $x$-direction as $B(x)$.
Consider an electron 
with the initial perpendicular energy $ E = m_e v_{\perp}^2/2$ at the cathode ($x=0$). 
The magnetic moment of the electron is conserved 
%if the electron experiences an almost constant magnetic 
%field relative to the gyro-frequency or  
if $\omega_{ce} \delta t > 1$, 
where $\delta t $ is the time scale with
 which the electrons experience a constant  magnetic field and  
$\omega_{ce} = eB(0)/m_e c $ is the gyro-frequency.   
Then, the conservation of the total kinetic energy of the electron is given as 
\begin{equation} 
\frac{m_e v_{x}^2}{2}  - e\phi(x) +  \frac{m_e v_{\perp}^2}{2} \frac{B(x)}{B(0)} = \mathrm{Const} \mathrm{,} \label{eq:pot}
\end{equation}
where the lasts term is derived from the conservation of the magnetic moment. 
Eq.~(\ref{eq:pot}) suggests that the electron would experience
 the additional potential from the magnetic mirror effect: 
\begin{equation}
 e\phi_{\mu} = - \frac{m_e v_{\perp}^2}{2} \left( \frac{ B(x)}{B(0)} -1 \right)
 \mathrm{.} \label{eq:mu}
\end{equation} 
Near the cathode, Eq.~(\ref{eq:mu}) can be recasted as 
\begin{equation}
 e\phi_{\mu} =  \frac{m_e v_{\perp}^2}{2}  k_{\mu} x
 \mathrm{,} \label{eq:mu2}
\end{equation} 
where $k_{\mu} =  (dB(0)/dx)/ B(0) $. 

\begin{figure}
\scalebox{0.6}{
\includegraphics{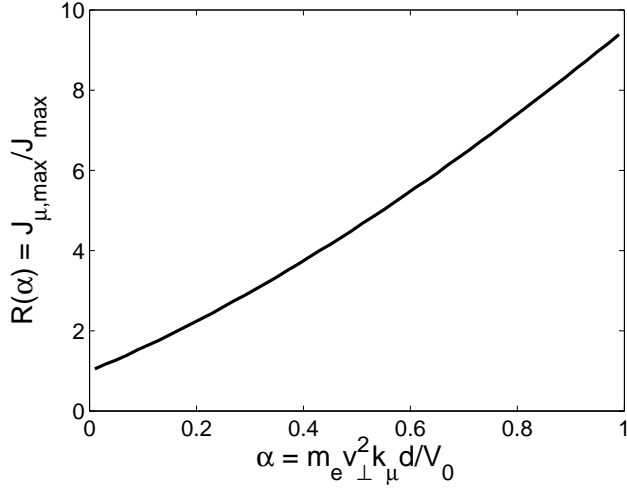}}
\caption{\label{fig:1}
The maximum current density in the presence of the additional potential
from the mirror effect, scaled by that in the absence of the additional
potential
%The ratios between the maximum current densities, in the presence and
%the absence of the additional  potential from the mirror effect,
($R(\alpha) = J_{\mathrm{\mu,max}}/ J_{\mathrm{max}}$;
see the discussion below Eq.~(\ref{eq:max}))
for a range of $\alpha = m_e v_{\perp}^2  k_{\mu} d / V_0$.
%Eq.~(\ref{eq:sta3}), together with the boundary condition $\phi(0) = 0 $ and $\phi(d) = V_0$,
%is integrated using the predictor-corrector method, then Eq.~(\ref{eq:max}) is
%used to determine the maximum current density.
We assume the magnetic field profile to be $B(x) = B(0) ( 1- k_{\mu} x) $
for $k_{\mu}x \leq 1$, and $B(x) = 0 $ for $k_{\mu}x > 1$.  
}
\end{figure}

The existence of an additional time-independent potential modifies the momentum and
Poisson equations to
\begin{eqnarray} 
  \frac{1}{2} m_e v_x^2(x) - e\left[\phi(x) +\phi_{\mu}(x)\right] = \mathrm{Const}  \mathrm{,} 
\nonumber \\ \nonumber \\ 
 \frac{ \partial^2 \phi(x) }{ \partial x^2  } =
\frac{4\pi J}{ \sqrt{ \frac{2e}{m_e} \left[\phi(x)+\phi_{\mu}(x)\right]}} \mathrm{.}  \label{eq:sta3} \\
 \nonumber 
\end{eqnarray}
The non-negativeness requirement becomes 
\begin{equation}
\phi(x) + \phi_{\mu}(x) > 0, \ \   \mathrm{for} \ \ 0 < x < d \mathrm{,} \label{eq:max}
\end{equation}
since $ \phi(x) + \phi_{\mu}(x) > \phi(0) + \phi_{\mu}(0)  =0$.  
%Note that the corresponding condition in the absence of the additional potential
 %from the mirror effect is $\phi(x) > 0 $.  
%The electrons can travel from the cathode to the anode at some point even when $\phi(x) < 0$,
%because only the inequality $\phi(x) + \phi_{\mu}(x) > 0$ needs to be satisfied.
Let us denote the maximum current achievable in the absence of the additional mirror potential 
by $J_{\mathrm{ max}}$ (see Eq.~(\ref{eq:child})),  
and the maximum current density in the presence of the mirror potential $\phi_{\mu}$
by $J_{\mathrm{\mu,max}}$.
In computing  $J_{\mathrm{\mu,max}}$,  Eq.~(\ref{eq:sta3})
is integrated for each value of $\alpha  = m_e v_{\perp}^2  k_{\mu} d / V_0$ from $x=0$ to $x=d$,
using the predictor-corrector method 
%such  that the solution satisfies   
with the boundary condition $\phi(0) = 0 $ and $\phi(d) = V_0$. 
The maximum current $J_{\mathrm{\mu,max}}(\alpha) $ is determined 
among the current densities that do not violate Eq.~(\ref{eq:max}). 
The ratio $R(\alpha) = J_{\mathrm{\mu,max}}/ J_{\mathrm{max}}$
increases with $ \alpha$ (Fig.~\ref{fig:1}), 
%increases with $ \alpha = m_e v_{\perp}^2  k_{\mu} d / V_0$ (Fig.~\ref{fig:1}), 

In the following, the regime of practical interest is estimated.  
As shown in Fig.~\ref{fig:1}, the critical parameter is 
\begin{equation}
\alpha =  \frac{m_e v_{\perp}^2}{V_0}   k_{\mu} d  \mathrm{.} \label{eq:alpha}
\end{equation} 
%This parameter measures the degree with which the Child Langmuir law could be overcome.
Let us consider the case where a microwave E\&M wave of the same frequency
as the cyclotron frequency ($\omega = e B(0)/ m_e c$)
injects the perpendicular kinetic energy to the electrons at the cathode. 
Assuming the electric field of the E\&M wave is perpendicular to 
the magnetic field,
the electron kinetic energy injected from the E\&M wave can be roughly estimated to be 
\begin{equation} 
v_{\perp}^2  \cong  \left( \frac{e E}{m_e \omega_{ce}} \right)^2 \left(\omega_{ce} \delta t_r \right)^2 \mathrm{,} \label{eq:v}
\end{equation} 
where $\delta t_r $ is the resonance interaction time of the electron with the E\&M wave and $E$ is the electric field of the E\&M wave. 
$\delta t_r $ can be controlled by changing the spot-size of the microwave source 
(for instance, if the E\&M wave is injected from the $y$-direction). However, 
$\delta t_r$  is limited due to the fact that the magnetic field varies spatially
while the frequency of the E\&M wave is fixed.
 We estimate $\alpha $ from Eqs.~(\ref{eq:alpha}) and (\ref{eq:v}) as
  $\alpha \cong ( e^2 E^2 / m_e \omega_{ce}^2 V_0) (\omega_{ce} \delta t_r ) $. 
If $\alpha > 1$, 
it is of plausible practical interest as shown in Fig.~\ref{fig:1}.

As an example, we consider an magnetic field of $B(0) =  1$ T.
$\alpha$ can be estimated to be
\begin{equation} 
\alpha \cong  0.2 \times 10^{-7} \frac{I}{V_0} (\omega_{ce} \delta t_r)^2  (k_{\mu} d) \mathrm{,}
\end{equation} 
where $V_0$ is in the unit of kV, and $I$ is the intensity of 
the microwave in the unit of $ \mathrm{J} / \mathrm{cm}^2 \sec $. 
%If $d = 10 \ \mathrm{cm}$ and $k_{\mu} = 1/\mathrm{cm} $, we estimate 
%$\alpha = 2 \times 10^{-7} (I/V_0)  (\omega_{ce} \delta t_r)^2$. 
If the cathode has the area of $s^2$ and the spot size of the microwave is the same with the cathode area, the condition $\alpha > 1$ can be recasted as 
\begin{equation} 
  P >  0.5 \times 10^{7} \frac{s^2 V_0}{(\omega_{ce} \delta t_r)^2 } \frac{1}{k_{\mu} d } \mathrm{,} \label{eq:P}
\end{equation}  
where $s$ is in the unit of $\mathrm{cm}^2$ and $P = s^2 I $ is the power of the microwave source.  
If $\omega_{ce} \delta t_r \cong 10 $, $d= 10 \ \mathrm{cm}$, $k_{\mu} = 1 / \mathrm{cm}$, $V_0 = 1 \ \mathrm{keV}$ and $s^2 = 1 \ \mathrm{cm}^2$,  Eq.~(\ref{eq:P}) is given as $P > 5 \  \mathrm{kJ}/ \sec $. 
If  $\omega_{ce} \delta t_r \cong 100 $, $d= 10 \ \mathrm{cm}$, $k_{\mu} = 0.1 / \mathrm{cm}$, $V_0 = 1 \ \mathrm{keV}$ and $s^2 = 1 \ \mathrm{cm}^2$  Eq.~(\ref{eq:P}) is given as $P > 500 \  \mathrm{J}/ \sec $.

To summarize, a scheme overcoming the Child-Langmuir law,
via the mirror effect of the magnetic moment conservation in the presence of a spatially varying magnetic field,
is discussed.
Due to the magnetic moment conservation, 
the spatially varying magnetic field generates an additional potential $\phi_{\mu}$ on the electrons and this potential can be designed in such a way that 
the self-induced potential $\phi$ is compromised in limiting the maximum current imposed from the Child-Langmuir law. 
Obtaining the optimal profile of the magnetic field $B(x)$  as a function of $x$, 
which  maximizes the achievable current for a given $v_{\perp}^2$  at the cathode, 
 is one interesting question.

\bibliography{tera2}% Produces the bibliography via BibTeX.

\begin{thebibliography}{8}
\expandafter\ifx\csname natexlab\endcsname\relax\def\natexlab#1{#1}\fi
\expandafter\ifx\csname bibnamefont\endcsname\relax
  \def\bibnamefont#1{#1}\fi
\expandafter\ifx\csname bibfnamefont\endcsname\relax
  \def\bibfnamefont#1{#1}\fi
\expandafter\ifx\csname citenamefont\endcsname\relax
  \def\citenamefont#1{#1}\fi
\expandafter\ifx\csname url\endcsname\relax
  \def\url#1{\texttt{#1}}\fi
\expandafter\ifx\csname urlprefix\endcsname\relax\def\urlprefix{URL }\fi
\providecommand{\bibinfo}[2]{#2}
\providecommand{\eprint}[2][]{\url{#2}}

\bibitem[{\citenamefont{Child}(1911)}]{child}
\bibinfo{author}{\bibfnamefont{C.~D.} \bibnamefont{Child}},
  \bibinfo{journal}{Phys.~Rev.} \textbf{\bibinfo{volume}{32}},
  \bibinfo{pages}{492} (\bibinfo{year}{1911}).

\bibitem[{\citenamefont{Langmuir}(1921)}]{lang}
\bibinfo{author}{\bibfnamefont{I.}~\bibnamefont{Langmuir}},
  \bibinfo{journal}{Science} \textbf{\bibinfo{volume}{54}},
  \bibinfo{pages}{(1386) 59} (\bibinfo{year}{1921}).

\bibitem[{\citenamefont{Ang et~al.}(2003)\citenamefont{Ang, Kwan, and
  Lau}}]{qchild}
\bibinfo{author}{\bibfnamefont{L.~K.} \bibnamefont{Ang}},
  \bibinfo{author}{\bibfnamefont{T.~J.~T.} \bibnamefont{Kwan}},
  \bibnamefont{and} \bibinfo{author}{\bibfnamefont{Y.~Y.} \bibnamefont{Lau}},
  \bibinfo{journal}{Phys.~Rev.~Lett.} \textbf{\bibinfo{volume}{91}},
  \bibinfo{pages}{208303} (\bibinfo{year}{2003}).

\bibitem[{\citenamefont{Lau et~al.}(1991)\citenamefont{Lau, Chernin, Colombant,
  and Ho}}]{qchild2}
\bibinfo{author}{\bibfnamefont{Y.~Y.} \bibnamefont{Lau}},
  \bibinfo{author}{\bibfnamefont{D.}~\bibnamefont{Chernin}},
  \bibinfo{author}{\bibfnamefont{D.~G.} \bibnamefont{Colombant}},
  \bibnamefont{and} \bibinfo{author}{\bibfnamefont{P.~T.} \bibnamefont{Ho}},
  \bibinfo{journal}{Phys.~Rev.~Lett.} \textbf{\bibinfo{volume}{66}},
  \bibinfo{pages}{1446} (\bibinfo{year}{1991}).

\bibitem[{\citenamefont{Griswold et~al.}(2012)\citenamefont{Griswold, Fisch,
  and Wurtele}}]{child3}
\bibinfo{author}{\bibfnamefont{M.~E.} \bibnamefont{Griswold}},
  \bibinfo{author}{\bibfnamefont{N.~J.} \bibnamefont{Fisch}}, \bibnamefont{and}
  \bibinfo{author}{\bibfnamefont{J.~S.} \bibnamefont{Wurtele}},
  \bibinfo{journal}{Phys.~Plasmas} \textbf{\bibinfo{volume}{19}},
  \bibinfo{pages}{024502} (\bibinfo{year}{2012}).

\bibitem[{\citenamefont{Griswold et~al.}(2010)\citenamefont{Griswold, Fisch,
  and Wurtele}}]{child4}
\bibinfo{author}{\bibfnamefont{M.~E.} \bibnamefont{Griswold}},
  \bibinfo{author}{\bibfnamefont{N.~J.} \bibnamefont{Fisch}}, \bibnamefont{and}
  \bibinfo{author}{\bibfnamefont{J.~S.} \bibnamefont{Wurtele}},
  \bibinfo{journal}{Phys.~Plasmas} \textbf{\bibinfo{volume}{17}},
  \bibinfo{pages}{114503} (\bibinfo{year}{2010}).

\bibitem[{\citenamefont{Northrop and Teller}(1960)}]{mirror}
\bibinfo{author}{\bibfnamefont{T.~G.} \bibnamefont{Northrop}} \bibnamefont{and}
  \bibinfo{author}{\bibfnamefont{E.}~\bibnamefont{Teller}},
  \bibinfo{journal}{Phys.~Rev.~} \textbf{\bibinfo{volume}{117}},
  \bibinfo{pages}{215} (\bibinfo{year}{1960}).

\bibitem[{\citenamefont{Keidar and Boyd}(2005)}]{mirror2}
\bibinfo{author}{\bibfnamefont{M.}~\bibnamefont{Keidar}} \bibnamefont{and}
  \bibinfo{author}{\bibfnamefont{I.~D.} \bibnamefont{Boyd}},
  \bibinfo{journal}{Appl.~Phys.~Lett.} \textbf{\bibinfo{volume}{87}},
  \bibinfo{pages}{121501} (\bibinfo{year}{2005}).

\end{thebibliography}

\end{document}